\documentclass[conference]{IEEEtran}

\usepackage{cmap}
\usepackage[utf8]{inputenc}
\usepackage[english]{babel}
\usepackage{hyperref}
\usepackage{indentfirst}
\usepackage{amsmath,amssymb,amscd,amsthm}
\usepackage{mathtools}
\usepackage{multirow}
\usepackage{multicol}

\usepackage{graphicx}
\usepackage{caption}
\usepackage{subcaption}
\usepackage{siunitx}
\usepackage{tikz}
\usepackage{cite}
\usepackage{url}
\usetikzlibrary{arrows}
\usetikzlibrary{shapes.multipart}
\usetikzlibrary{plotmarks}
\usetikzlibrary{patterns}

\newcommand\copyrighttext{%
\footnotesize \textcopyright \enspace 2019 IEEE. Personal use of this material is permitted. Permission from IEEE must be obtained for all other uses, in any current or future media, including reprinting/republishing this material for advertising or promo
tional purposes, creating new collective works, for resale or redistribution to servers or lists, or reuse of any copyrighted component of this work in other works.  DOI: \href{https://doi.org/10.1109/BlackSeaCom.2019.8812785}{10.1109/BlackSeaCom.2019.8812785}
}
\newcommand\copyrightnotice{%
\begin{tikzpicture}[remember picture,overlay]
\node[anchor=south] at (current page.south) {\fbox{\parbox{\dimexpr\textwidth-\fboxsep-\fboxrule\relax}{\copyrighttext}}};
\end{tikzpicture}%
}

\begin{document}

\title{IEEE 802.11ba --- Extremely Low Power Wi-Fi \\ for Massive Internet of Things --- \\ Challenges, Open Issues, Performance Evaluation\thanks{The research was supported by RFBR grant 18-07-01356 a.}} 

\author{\IEEEauthorblockN{
		Dmitry Bankov\IEEEauthorrefmark{1}\IEEEauthorrefmark{3}, 
		Evgeny Khorov\IEEEauthorrefmark{1}\IEEEauthorrefmark{3}, 
		Andrey Lyakhov\IEEEauthorrefmark{1},
		Ekaterina Stepanova\IEEEauthorrefmark{1}\IEEEauthorrefmark{3}
	}
	\IEEEauthorblockA{ \IEEEauthorrefmark{1}Institute for Information Transmission Problems, Russian Academy of Sciences, Moscow, Russia\\
		\IEEEauthorrefmark{3}Moscow Institute of Physics and Technology, Moscow, Russia \\    
		Email: \{bankov, khorov, lyakhov, stepanova\}@iitp.ru}
}

\maketitle
\copyrightnotice

\begin{abstract}
	Many recent activities of IEEE 802.11 Working group have been focused on improving power efficiency of Wi-Fi to make it favorable for massive Internet of Things scenarios, in which swarms of battery supplied sensors rarely communicate with remote servers. The latest step towards this direction is the work on a new IEEE 802.11ba amendment to the Wi-Fi standard, which introduces Wake-Up Radio. This radio is an additional interface with extremely low power consumption that is used to transmit control information from the access point to stations while their primary radio is switched off.
	This paper describes the IEEE 802.11ba protocol, discusses its open issues, investigates several approaches to provide energy efficient data transmission with 802.11ba, and evaluates how much 802.11ba improves energy efficiency and even reduces channel time consumption.
\end{abstract}

\section{Introduction}
\label{sec:intro}

Wi-Fi shows an outstanding success story, having become more than just a technology for home and office networks. Recent activities of the Wi-Fi community are focused on adapting this technology to the growing market of the Internet of Things (IoT) with its tremendous number of autonomous devices, most of which are battery supplied. Originally designed to provide connectivity for a small number of computers and laptops that transmit relatively heavy flows, Wi-Fi is not suitable for such use cases. Two main challenges are (i) to provide connectivity to a large number of devices connected to a single access point (AP), and (ii) to make communications of IoT devices energy efficient. The latter is especially important since these devices need to live with a single battery for years. 

A significant step towards addressing these challenges has been done with the 802.11ah aka Wi-Fi HaLow \cite{khorov2015survey}.
One of its novelties is the Target Wake Time (TWT) that allows the AP to negotiate the schedule of frame exchange with a station (STA) in advance.
TWT enables the STA to switch off its radio until the negotiated time and thus not to spend energy on unnecessary listening to the channel between the scheduled transmissions.
In contrast to the basic Wi-Fi power management, the STA does not even have to wake up periodically to receive beacons.
Scheduling also reduces contention by spreading the transmissions of various STAs over time.
Thus, TWT reduces the time a STA needs to be active and saves much energy for those STAs, which rarely transmit data \cite{khorov2015survey}. 

Being very promising for IoT, 802.11ah would master the market but for two issues. First, it operates in sub \SI{1}{\GHz} and, thus, is not backward compatible with legacy networks.
The second reason is that the appearance of 802.11ah was overlapped with the active work on 802.11ax which is the successor of 802.11ac as the mainstream Wi-Fi.
So, key Wi-Fi vendors shifted their efforts to 802.11ax.
Being designed for dense deployment, 802.11ax introduces new channel access methods reducing contention and adapts TWT from 802.11ah \cite{khorov2018tutorial}.
This and the fact that it uses traditional \SI{2.4}{\GHz} and \SI{5}{\GHz} make it a promising solution not only for smartphones and laptops but also for IoT scenarios.

Whatever technology --- 802.11ah or 802.11ax --- is used, TWT has a significant drawback. Designed to avoid carrier sense during long time intervals, TWT suffers from clock inaccuracy.
Specifically, the Wi-Fi standard allows clock drifting of up to $100$ ppm, i.e., the clock drift may reach \SI{0.36}{\s} per hour.
It means that if a STA is scheduled to wake up in an hour, it shall wake up \SI{0.36}{\s} in advance, listen to the channel till it receives a beacon with the value of AP's clock, or contend for the channel with other STAs to indicate its awake state to the AP.
In both cases, the STA needs to be awake much more time than it is really needed to send or receive a packet.

To reduce the energy consumption caused by unnecessary listening to the channel, the IEEE 802 LAN/MAN committee is developing the 802.11ba amendment \cite{802.11ba}.
It introduces an auxiliary very simple radio, called Wake-Up Radio (WUR).
WUR consumes thousands of times less than the Primary Communication Radio (PCR) and is used to receive special wake-up frames or synchronization information sent by the AP to the STAs.
In this paper, we analyze this technology, discuss its potential open issues and describe a way to effectively use it to reduce energy consumption in a heterogeneous network, where apart from a large number of IoT devices with light traffic, there are several laptops which generate heavy flows.

The rest of the paper is organized as follows.
Section \ref{sec:power} briefly introduces Wi-Fi power management framework. In Section \ref{sec:description}, we describe new features of 802.11ba, discuss how to use them and summarize open issues. In Section~\ref{sec:results}, we compare the performance of 802.11ba with existing 802.11ax methods and show that 802.11ba can manifold reduce power consumption.
Section \ref{sec:conclusion} concludes the paper.

\section{Power Management in Legacy Wi-Fi}
\label{sec:power}

\textbf{Basic Power Management Approach.} Since its first version, Wi-Fi contains a power management framework, which defines two modes of operation. In the active mode, the STA is always awake and can transmit and receive frames. In the power save (PS) mode, it alternates between two states: awake and doze. In the doze state, the STA switches off its radio, and can neither transmit nor receive. 

In infrastructure networks, a STA shall notify the AP before changing the mode of operation.
If the STA is in the PS mode, the AP buffers all frames destined for this STA. To notify PS STAs about the buffered packets, the AP includes in each beacon a Traffic Indication Map (TIM) which indicates the presence of packets destined for each STA. Periodically, beacons contain a Delivery Traffic Indication Map (DTIM) element that notifies whether the AP has buffered groupcast packets. The AP broadcasts groupcast packets right after a beacon with DTIM.
Every PS STA periodically wakes up to receive beacons. The STA does not need to listen to each TIM element. In practice, PS STAs wake up before DTIM beacon. If no buffered packets are destined for the STA, it returns to the doze state right after the beacon. Otherwise, the STA sends a PS-Poll frame after groupcast transmission. As a response to the PS-Poll, the AP sends buffered frames.

To transmit a frame, the STA does not need to wake for a beacon. Instead, it shall switch on its radio, wait for any frame reception (but no longer than the Probe Delay timeout), and only after that start accessing the channel. Such a defer is caused by the peculiarities of Wi-Fi channel access.

To access the channel, Wi-Fi devices use Enhanced Distributed Channel Access (EDCA), which is a sort of CSMA/CA with truncated binary exponential backoff. In addition to physical carrier sense, Wi-Fi devices use virtual carrier sense, called the Network Allocation Vector (NAV), which uses the Duration field in frame headers. This field indicates how long the channel will be virtually busy after the end of the frame.
A STA that has been in the doze state could miss a frame setting the NAV. Thus, it needs to receive another frame to understand if the channel is occupied virtually or not. 

During almost 30-year Wi-Fi development process many power saving mechanisms have been designed.
For example, with Unscheduled Automatic Power Save Delivery (U-APSD), a STA can retrieve the buffered packets from the AP not only after a beacon but any time when it transmits a packet in uplink.
However, most of the power management methods require periodic beacon reception, which consumes much energy in comparison to frame transmission once per hour or less often.

\textbf{TWT.} The most powerful power saving mechanism designed by today in Wi-Fi is TWT. While a detailed description of TWT in 802.11ah and 802.11ax can be found in \cite{khorov2015survey,khorov2018tutorial}, in this paper we briefly summarize its main peculiarities.
TWT allows a STA to negotiate with the AP time instants when the STA wakes up for some time (called TWT Service Period, TWT SPs) and exchanges frames with the AP.
With TWT, the STA can stay doze always except for the negotiated TWT SPs and is not required anymore to wake up for beacons, which reduces energy consumption significantly.
If a STA needs to retrieve data from the AP, it sends a PS-Poll frame. If the AP wants to trigger uplink transmission, it can send a trigger frame (TF) defined in 802.11ax.

Note that an established TWT SP does not protect TWT transmission from collisions with other STAs since it does not forbid other STAs to access the channel. To protect TWT transmissions, NAV can be used. For example, the AP can send a CTS-to-self frame before a TWT transmission.

The main drawback of TWT is that the STA loses time synchronization with the AP during long doze state because of clock drift.
The standard allows clock drifting of up to 100 ppm.
For this reason, in case of rare traffic, the STA needs to wake up well in advance before the negotiated TWT SP, and either to wait until receiving a frame from the AP or to contend for the channel in order to transmit its frame.
As shown in \cite{stepanova2018clock}, clock drifting significantly reduces the efficiency of TWT.
A possible way to save energy is to forbid other STAs to transmit during the TWT SP plus two guard intervals, each of which covers the possible clock drifts.
With such a method, the STA can transmit as soon as it wakes up without contention.
To reduce channel time consumption, TWT of several STAs can be located close to each other.
But the reserved time is enormous.
Moreover, the standard does not allow reserving such a long time interval. 
The described drawbacks and inefficiency of existing methods have motivated the further improvement of power saving techniques with the use of addition low-power wake-up radio.
This is done with 802.11ba.
Its current version is D2.0, while the final version of the standard is scheduled at the end of 2020.

\section{IEEE 802.11ba}
\label{sec:description}

The main feature of IEEE 802.11ba is a simple low-power WUR that can be used to wake up a STA exactly in a given time.
WUR is only used while PCR is off.
WUR is not designed for user data transmission, but it is used to transmit management information only from the AP to the STAs, such as wake-up notifications.
This explains why in 802.11ba the STAs have only a WUR receiver, but not a WUR transmitter.

Naturally, the concept of WUR is not brand-new. Many low-power radios have been designed to achieve power consumption as low as 1~mW \cite{pletcher20082ghz, hambeck20112, salazar201513}. Moreover, before 802.11ba several WUR-related MAC protocols \cite{sthapit2011effects, al2011mac} have been created.

The main challenge that was addressed while developing 802.11ba is how to make WUR compatible with legacy Wi-Fi in the following senses. First, the AP shall be able to transmit WUR signals in the same bands where traditional Wi-Fi devices work. Second, the rules of WUR operation shall take into account Wi-Fi peculiarities. Below, we describe the main features of 802.11ba focusing on the open issues.

\begin{figure}[!tbp]
	\centering
	\includegraphics[width=0.9\linewidth]{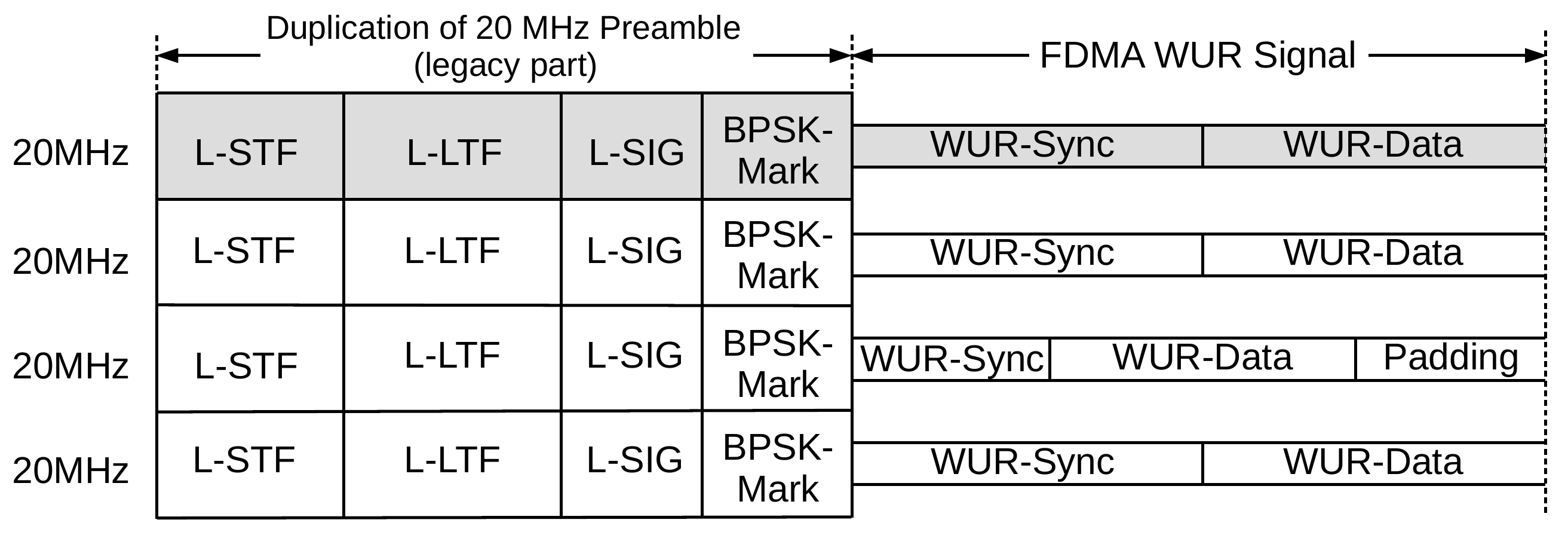}
	\caption{FDMA transmission of WUR frames in 80 MHz }
	\label{fig:}
	\vspace{-1.5em}
\end{figure}

\textbf{PHY.} While the AP transmits a WUR frame, legacy devices that operate in the same band shall understand that the channel is busy. For that, every WUR transmission starts with a legacy Wi-Fi preamble (see Fig.~\ref{fig:}, a single 20~MHz band transmission is highlighted with gray) transmitted in 20~MHz bands. The preamble consists of training fields (namely, L-STF and L-LTF), and the SIGNAL field (L-SIG) that determines the frame duration. Legacy devices capture and decode the preamble and consider the channel as physically busy for the whole WUR frame duration. The preamble is followed by a BPSK-Mark field, which is added to prevent 802.11n devices from switching to the channel idle state.

The rest of the WUR frame can be only received by WUR. To reduce power consumption for the receiver, a very simple on-off keying (OOK) modulation and a narrow band of 4~MHz.
This part of the WUR frame consists of two fields: WUR-Sync and WUR-Data.

When PCR is off, the STAs cannot capture the legacy preamble.
To allow WUR receivers to synchronize to a WUR frame, WUR-Sync is used. Apart from that, the structure of WUR-Sync determines which of the two rates: low data rate (LDR) and high data rate (HDR) is used to transmit WUR-Data.
Specifically, to indicate HDR, WUR-Sync contains a unique 32-bit sequence, each bit being represented by a \SI{2}{\us} OOK symbol. For LDR, the sequence is inverted and duplicated. 
WUR-Data carries a payload from the MAC layer encoded with a Manchester-based code. In case of LDR, input bit $1$ is encoded as $1010$ and input bit $0$ is encoded as $0101$ which are transmitted with \SI{4}{\us} OOK symbols. 
In case of HDR, input bit $1$ is encoded as $10$ and input bit $0$ is encoded as $01$ which are transmitted with \SI{2}{\us} OOK symbols. Thus, WUR-Data can be transmitted at either 62.5 kbps or 250 kbps. 

\textbf{MAC.}
IEEE 802.11ba defines four types of MAC frames: (i) wake-up frames that trigger the STA to switch on its PCR, (ii) WUR beacons that are periodically sent to provide timing synchronization, (iii) WUR discovery frames that are used to allow the STA for low power network discovery without interruption of the connectivity with the current AP, and (iv) vendor-specific frames that are out of the scope of the standard. At the MAC layer, WUR frames are rather short and contain: (i) the Frame Control field (8 bits) that determines the type of the frame and its length if the length exceeds 48 bits, (ii) shortened address (12 bits), (iii) type dependent control information (12 bits), e.g., the current clock value, the counter indicating crucial network configuration updates, etc., (iv) Optional Frame Body, and (v) frame checksum (16 bits).

WUR frames are transmitted according to the EDCA rules.
Since the STAs do not use WUR for transmission, even unicast WUR frames are sent without acknowledgment.
Consequently, the retry counter and the contention window of the binary exponential backoff are not increased if the transmission fails.

\textbf{Wake-up Procedure.}
To enable WUR, the STA shall negotiate its parameters with the AP. Specifically, the devices agree on a channel that is used for WUR frames transmission, the power management method which is used to retrieve the data from the AP after the STA is woken up. For that, the basic power management approach, U-APSD, TWT, or another method can be used. After such a negotiation, the STA can switch off its PCR and switch on the WUR.

To notify a STA to switch on its PCR (e.g., if the AP has buffered data for this STA), the AP sends a WUR wake-up frame. Switching PCR on takes some time the maximal value of which is indicated by the STAs to the AP. To notify the AP about waking up, the STA may send some frame to the AP. It can be an uplink data frame or a frame which retrieves the buffered data, e.g., a PS-Poll. The AP can only transmit data in the downlink with PCR if it receives such a frame or the switching on timeout indicated by the STA has expired. 

A Wake-up frame can be unicast or groupcast. For example, \cite{tang2017energy} evaluates delays induced by WUR and proposes to periodically broadcast Wake-Up frames to wake up all WUR STAs. Such an approach decreases the number of WUR frames, but increases contention and, thus, energy consumption.

\textbf{WUR Duty Cycle.}
Thanks to WUR, the PCR can be in the doze state all the time, except for the time, when it is used for data transmission or reception. However, when PCR is off, WUR is on and consumes energy, though its power consumption is much smaller than that of PCR.
To address this issue, 802.11ba introduces a duty cycle mode. With this mode, the AP and the STA agree on the strictly periodic time intervals during which WUR shall be switched on. During the remaining time, both WUR and PCR may be switched off. WUR Duty Cycle allows the AP to reduce the time that the STA's WUR is in the awake state and to separate the activity of WUR interfaces of multiple STAs over time. The price for this is the increased delay for downlink transmission.

\textbf{WUR Beacons.}
If the STA rarely transmits or receives data, the clock drift may reduce power efficiency, since the STA may switch on its radio(s) too far from the scheduled time. To maintain connectivity with the AP and to maintain time synchronizations, when the STA’s PCR is off, the AP may periodically broadcast WUR beacons. Specifically, WUR beacons contain the partial timestamp of the current clock value. They are sent in a way similar to traditional ones, however with a longer period, to reduce overhead.

\textbf{Channellization.} 
Since a WUR frame is long, it is necessary to group several wake-up frames. For example, the authors of \cite{hong2018low} propose and study a scheme of grouped wake-ups in an 802.11ax network which enables uplink OFDMA. In the proposed scheme, a wake-up packet is followed by a TF of 802.11ax which allocates resources for uplink transmission. 

The usage of groupcast wake-ups is non-flexible and non-efficient with a large number of groups or if re-grouping of STAs is continuously needed.
To provide a more robust solution, 802.11ba allows transmitting several WUR frames in parallel at different frequencies (i.e., with FDMA), one per 20~MHz subchannel, if the network operates in a $\ge$~20~MHz (e.g., 80~MHz) band.
In this case, the legacy preamble is duplicated in every 20~MHz subchannel, while the WUR parts are different.
Note that although WUR transmission occupies only 4~MHz, multiplexing several WUR frames in the same 20~MHz channel is forbidden to simplify the receiver. 
The WUR frames are aligned in time using padding (see Fig.~\ref{fig:}).
It is done to ensure that a STA receiving a WUR frame cannot start an uplink transmission with PCR while in the primary channel WUR transmission is still not finished.
Following Wi-Fi channel bonding rules of 802.11ax, the described FDMA transmission shall include the primary subchannel of the network, while all secondary subchannels can be punctured if busy.
When negotiating WUR operation, the AP and the STA choose the subchannel in which WUR frames destined for this STA will be transmitted.

\textbf{Open Issues.}
The design of the WUR operation raises many issues.
First, how to deal with different coverage and reliability of traditional signals and WUR signals?
Second, it is not clear how WUR transmissions interfere with legacy ones.
Third, what will happen in case of collision of a long WUR frame with a short legacy one?
Specifically, will the legacy devices detect the channel as busy after the legacy frame?
Fourth, which power management method shall be used with WUR?
It seems that the correct answer depends on the traffic pattern and rate.
Fifth, how to group STAs to wake them up in order to reduce channel time consumption without significant losses in energy efficiency?
Sixth, which device, the AP or the STA, shall transmit the first frame after the wake-up frame?
On one side, if the STA wakes up and immediately contends for the channel, it may transmit earlier, which means lower energy consumption.
On the other side, by using TFs followed by uplink OFDMA transmission of 802.11ax, we can increase the efficiency of the channel usage.
Seventh, what shall be WUR beacon period?
Power consumption analysis shows that in case of duty cycle operation it is not worth to wake up WUR only for time synchronization.
If so, is it better to use aperiodic beacons sent only during the time intervals when WURs of several STAs are on?
Eighth, how to map WUR subchannels to various STAs, taking into account that if a secondary channel is busy, it is punctured and wake-up frames are not sent?

The answers to these questions have a substantial impact on the performance of WUR and require in-depth research.
We further evaluate the gain provided by 802.11ba in comparison to the 802.11ax power saving methods. 

\section{Performance Evaluation of WUR}
\label{sec:results}
\begin{figure}[t]
	\centering
	\center{\includegraphics[width=0.9\linewidth]{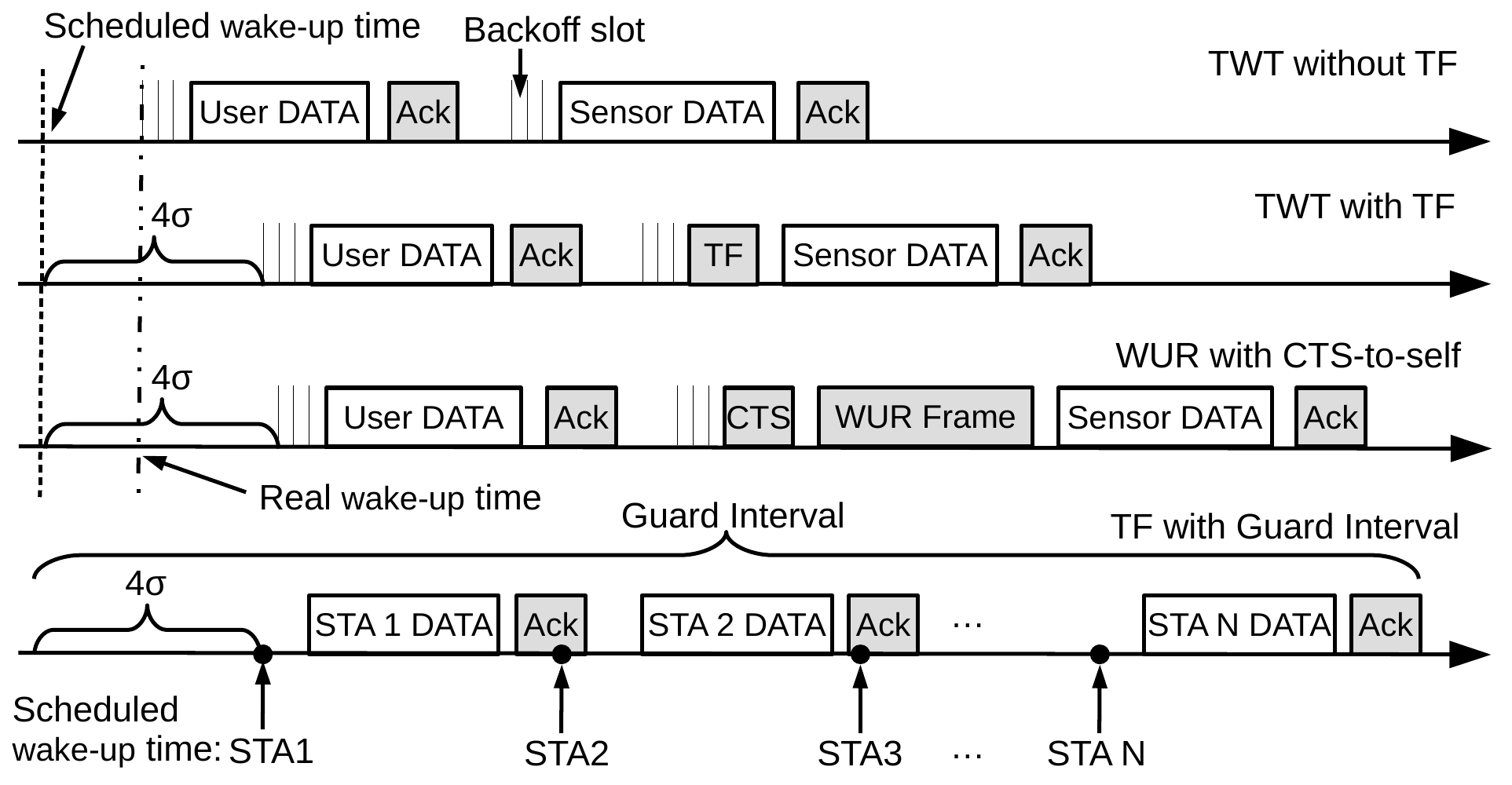}}
	\caption{Considered transmission methods}
	\label{fig:cases}
	\vspace{-1.5em}
\end{figure}

Consider an 802.11ax network with one AP, $M$ usual STAs that generate saturated traffic and $N$ sensor STAs, each of which rarely transmits a data frame to the AP.
Because of space limitation we consider only uplink transmission, while the downlink can be evaluated similarly. 
All the STAs are in the transmission range of each other.
The STAs use TWT to reduce power consumption.
The AP schedules time for each sensor STA when it will transmit its uplink frame.
So the sensors switch off their PCR until the target time.
At a scheduled time, each sensor wakes up (independently from each other) and starts the transmission.
Because of clock drifting, the awakening time differs from the target time by a random shift, normally distributed with variance $\sigma^2$.
We consider four methods (see Fig.~\ref{fig:cases}) to organize the transmission.

\textbf{1. TWT without TFs.} With this method, a sensor wakes up at a scheduled time and transmits its data with EDCA. Thus, it contends for the channel with the STAs with saturated traffic.

\textbf{2. TWT with TFs.} With this method, a sensor wakes up at a scheduled time and waits for a TF. The AP generates the TF $4 \cdot \sigma$ after the scheduled time to guarantee that the probability of the STA missing the TF because of the clock drift is less than 0.01\%. To send the TF, the AP contends for the channel with the usual STAs. After receiving the TF, the sensor responds to the AP with a data frame sent without contention.

\textbf{3. TWT with guard interval.} With this method, the AP eliminates contention between sensors and the usual STAs as follows. It groups sensors and allocates sensors' wake-up times one by one with the distance between neighboring times twice longer than a successful data frame transmission, as recommended in \cite{stepanova2018clock}. The AP forbids other STAs to transmit during the time interval that starts $4 \cdot \sigma$ before the first scheduled time and ends with the last sensor transmission. 

\textbf{4. WUR with CTS-to-self.} With this method, the network operates according to 802.11ba with duty cycles. Duty cycles of various STAs are individually assigned to avoid intersection of scheduled transmissions. As in the second method, the AP starts accessing the channel $4 \cdot \sigma$ after the scheduled time. Then it transmits a CTS-to-self frame which informs the surrounding STAs that the channel will be busy during the interval needed to receive the data from the sensor. After the CTS-to-self transmission, the AP uses its WUR to send a Wake-Up frame to the sensor. When the sensor receives the WUR frame, it switches on its PCR and transmits the data frame.

To compare the efficiency of these methods, we consider the energy consumed by the sensor STAs to transmit one data frame measured from the wake-up time until the frame delivery, including the energy required to listen to the channel. We also estimate the amount of channel time consumed by sensors. This time includes the collisions involving the AP or the sensor STAs and the time reserved for transmissions.

To evaluate the efficiency of the considered methods we implement them in the ns-3 network simulator. In our experiments, the network operates on a 20~MHz channel and consists of 10 usual STAs and 10 sensor STAs. The STAs use the most reliable modulation and coding scheme MCS0, which results in the duration of the data transmissions of \SI{1480}{\us}. The WUR frames are transmitted with LDR, and their duration is \SI{920}{\us}. 

\begin{figure}[t]
	\center{\includegraphics[width=0.8\linewidth]{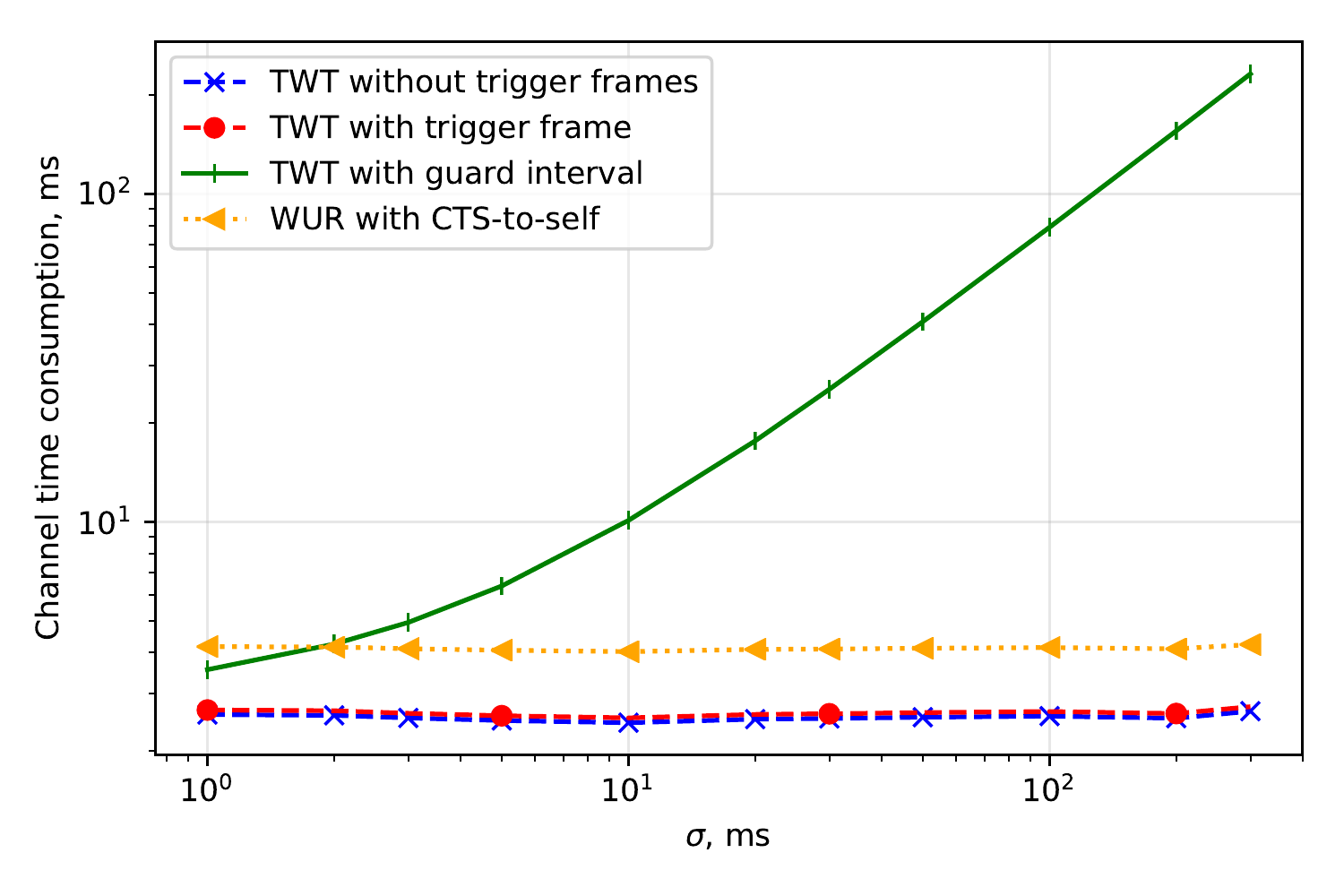}}
	\vspace{-1em}
	\center{\includegraphics[width=0.8\linewidth]{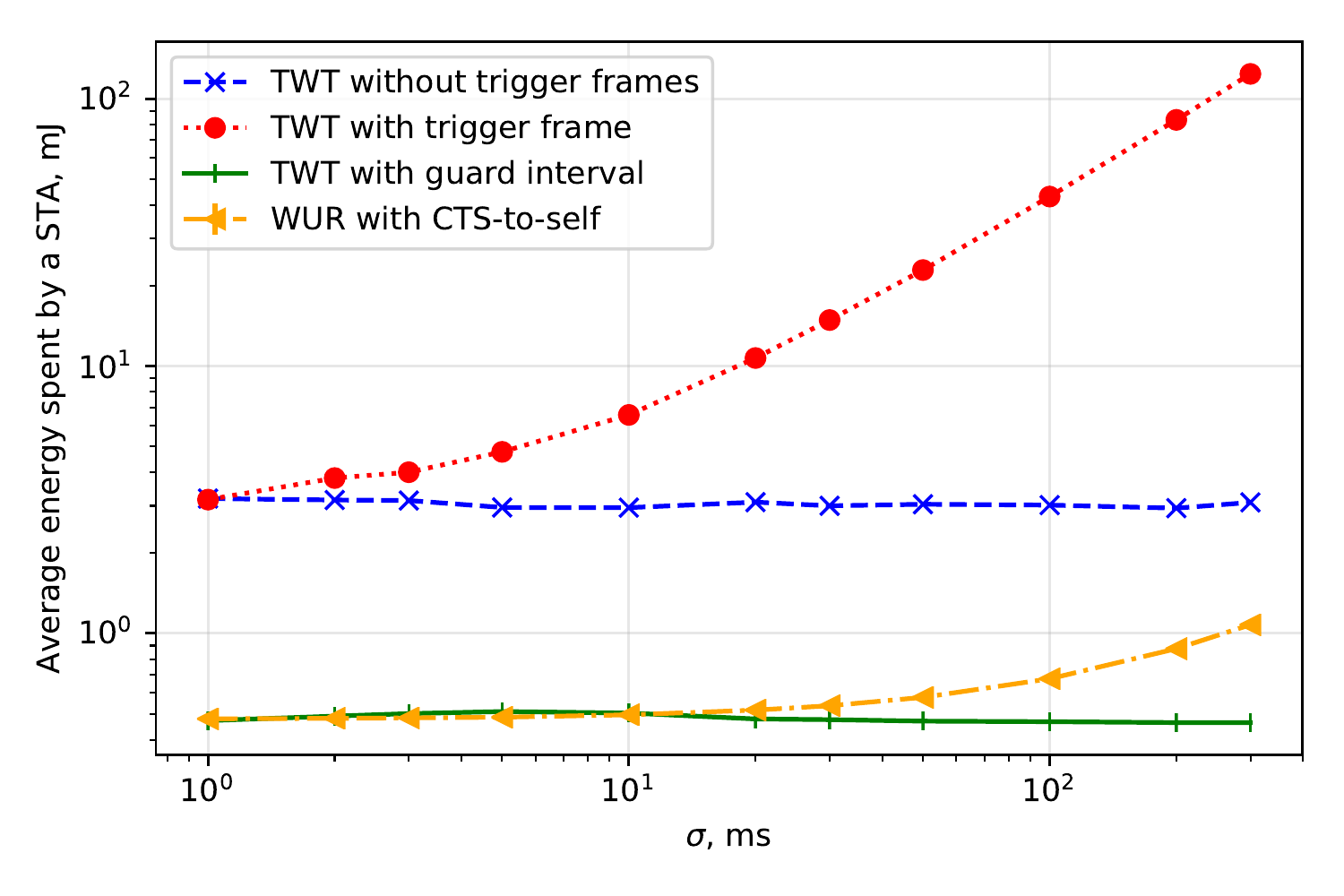}}\vspace{-1em}
	\caption{Channel time and energy consumption vs. clock drift}
	\label{ris:time_all}
	\vspace{-1.5em}
\end{figure}

Fig.~\ref{ris:time_all} shows the dependency of the channel time and energy consumption per frame on the variance of the clock drift.
The channel time depends on the variance only for TWT with guard interval because the guard interval grows with $\sigma$.
Note that because of STA grouping, the long guard intervals are shared between several transmissions and the channel time consumption per frame is even less than a guard interval.
The more STAs are grouped, the higher is channel time saving caused by grouping.
Nevertheless, the absolute value is too high in comparison to other methods, for which the main factor of the channel time is the total duration of the frames required to be sent to deliver data.
Specifically, for WUR, channel time consumption is 50\% higher than for TWT without guard interval because of long WUR frame transmission. 

As for energy consumption, the use of guard interval or WUR manifold reduces energy consumption with respect to pure TWT with/without TF. TF even worsens performance in case of high clock drifting, since the sensor STAs need to listen to many frames while waiting for the TF. WUR can provide the smallest energy consumption provided that the clock drift is not too high. For a high clock drift ($\sigma=100$ ms), its performance degrades since the WUR receiver spends energy on channel listening. Since its power consumption is less than that of PCR, the total energy consumption increase is not so dramatic as in the case of TWT with TF. To improve performance in case of even higher $\sigma$, the AP should send WUR beacons more often when the STA's WUR is probably awake. Having received the timestamps, the sensor STA resumes its time synchronization, corrects the clock drift and can switch off its WUR till the correct scheduled time.

Although, TWT with guard interval provides smaller power consumption than WUR, in case of high $\sigma$. However, for that much more channel time shall be reserved. In practice, reserving dozens or even hundreds of milliseconds of the channel time is impossible. Thus, in spite of results, this method cannot be used in real devices for high $\sigma$. Taking this into account, we can conclude that 802.11ba can provide high performance in both metrics: energy consumption, and the portion of occupied channel time. 

\section{Conclusion}
\label{sec:conclusion}

In the paper, we have observed and discussed the main features of the novel 802.11ba technology, which focuses on significant reduction of energy consumption.
We have shown how to use these features together and evaluated the impact of the low-power WUR on power and channel time consumption in a heterogeneous network with both battery-supplied sensor STAs rarely transmitting or receiving data and usual STAs with heavy traffic.
Besides, we have found many other issues which significantly affect WUR performance.
We have pointed out such open problems and provided some ideas on their solution. 

\bibliographystyle{IEEEtran}
\bibliography{biblio}

\end{document}